\documentclass[12pt]{amsart}

\usepackage{amsmath}
\usepackage{latexsym}
\usepackage{amsfonts}
\usepackage{amssymb}

\newtheorem{lemma}{Lemma}

\theoremstyle{definition}
\newtheorem{remark}{Remark}
\newtheorem{example}{Example}

\newcommand{\mh}{\mathcal{H}}
\newcommand{\mk}{\mathcal{K}}
\newcommand{\mm}{\mathcal{M}}

\newcommand{\R}{\mathbb R}
\newcommand{\C}{\mathbb C}
\newcommand{\N}{\mathbb N}
\newcommand{\bor}[1]{\mathcal{B}({#1})}
\newcommand{\br}{\mathcal B(\mathbb R)}

\newcommand{\ip}[2]{\left\langle\,#1\,|\,#2\,\right\rangle}
\newcommand{\ket}[1]{\mid#1\rangle}
\newcommand{\kb}[2]{|#1\,\rangle\langle\,#2|}
\newcommand{\no}[1]{\parallel#1\parallel}

\begin{document}

\title[Noise and disturbance]{Noise and disturbance in quantum measurement}

\author[Busch]{Paul Busch}
\address{Paul Busch, Department of Mathematics, University of Hull, Hull HU6 7RX}
\email{P.Busch@hull.ac.uk}

\author[Heinonen]{Teiko Heinonen}
\address{Teiko Heinonen, Department of Physics and Department of Mathematics, University of Turku, FIN-20014 Turku, Finland}
\email{teiko.heinonen@utu.fi}

\author[Lahti]{Pekka Lahti}
\address{Pekka Lahti, Department of Physics, University of Turku, FIN-20014 Turku, Finland}
\email{pekka.lahti@utu.fi}

\maketitle{}

\begin{abstract}
The operational meaning of some measures of noise and
disturbance in measurements is analyzed and their limitations are
pointed out. The cases of minimal noise and least disturbance are
characterized.
\end{abstract}

\section{Introduction}

No physical measurement is absolutely accurate. It seems inevitable that there
will always be a residual degree of uncertainty as to how close the outcome is
to what should have been expected. Likewise, a measurement, being an
interaction of the apparatus with the measured system, must always be expected
to effect some change, or disturbance, of the measured system. In classical
physics it seems possible to achieve arbitrary levels of accuracy and to make
the disturbance as small as one wishes. These options appear to be ruled out in
quantum physics, due to the fact that there are pairs of physical quantities
which cannot be measured together. Such quantities are represented by mutually
noncommuting operators or operator measures.

In his fundamental work of 1927 on the interpretation of quantum mechanics,
W.~Heisenberg sketched two versions of what became known as the uncertainty
principle and which can be vaguely summarized as follows:
\begin{itemize}
\item[(UP1)] A measurement, with inaccuracy $\epsilon(A)$, of a quantity $A$ that does not
commute with a quantity $B$ will disturb the value of $B$ by an
amount $\eta(B)$ such that an appropriate pay-off relation holds
between $\epsilon(A)$ and $\eta(B)$.
\item[(UP2)] A joint measurement of two noncommuting quantities $A,\ B$ must be imprecise,
with the inaccuracies $\epsilon(A)$, $\epsilon(B)$ satisfying an
uncertainty relation.
\end{itemize}

Heisenberg focussed on pairs of canonically conjugate observables
and he gave model experiments to demonstrate that relations of the
form $\epsilon(A)\eta(B)~\sim h$ and $\epsilon(A)\epsilon(B) \sim h$
had to hold in the cases (UP1) and  (UP2), respectively.

The quantities $\epsilon,\ \eta$ were not formally or operationally defined but
simply intuitively identified with measures of the spread of wave functions or
momentum amplitudes. It took several decades of research into quantum
measurement theory until concepts of imprecise and joint measurements of
noncommuting quantities were developed, with an appropriate definition of
measures of inaccuracy and disturbance that allowed one to give rigorous
formulations of the uncertainty principle in its versions (UP1) and (UP2) for
conjugate quantities. A review of the theory of joint measurements leading to
(UP2) in the case of position and momentum can be found in \cite{OQP}. A
formalization of (UP1) and conditions for its validity have been obtained in
recent years by M.~Ozawa \cite{Oza1, Oza3, Oza4}, see also his preprint
\cite{Oza5}.

In this paper we study the measures of measurement imprecision, or
\emph{measurement noise}, and \emph{disturbance} used in these investigations.
On closer inspection it turns out that these quantities do not satisfy some
requirements that one might reasonably expect of measures of measurement noise
and disturbance. Moreover, their definitions do not seem to apply to more
general types of measurement where the observables intended to be measured are
represented by positive operator measures which are not projection valued and
which may even be noncommutative. We will highlight some of the shortcomings of
these notions and consider possible ways of finding more suitable measures.

\section{Measurement Noise}\label{mnoise}

The intuitive idea of noise in a measurement can be captured as
the dissimilarity between the actually measured probability
distribution and the distribution of the observable intended to be
measured. In quantum mechanics these probability measures are
determined by positive operator measures: one, $E^{\mathcal{M}}$,
that represents the quantity that is actually measured by a given
measurement process $\mathcal{M}$, and another one, $E$, that
represents the observable intended to be measured. We will usually
assume that the operator measures are bounded so that their moment
operators are bounded and selfadjoint.

Any quantity describing measurement noise could be expected to
have the following properties. First, it should be possible to
estimate the noise by comparing the statistics of the measurement
in question with the statistics of a `good' measurement of the
quantity in question (provided that such a `good' measurement is
available for the purpose of calibration of the new measurement).
This means that the noise quantity should be a function
$\epsilon(E,E^{\mathcal{M}},\psi)$  of the input state $\psi$ and
the two observables involved. Second, whenever the noise is
`small', this should mean that the measurement is `good'. We take
this to mean that vanishing noise (in a given state $\psi$) should
indicate that the probability distributions $E_\psi$ and
$E^{\mathcal{M}}_\psi$ of $E$ and $E^{\mathcal{M}}$ are the same
in that state. Finally, if the measurement is a good one, meaning
that $E_\psi=E^{\mathcal{M}}_\psi$ for all states $\psi$, then
this should be indicated by a vanishing noise measure for all
$\psi$.

A noise measure that satisfies all these requirements is given by the total
variation norm of the difference between the probability measures,
$\epsilon_1(E,E^{\mathcal{M}},\psi)=\|E_\psi-E^{\mathcal{M}}_\psi\|_1$, see
Section \ref{total}. Other frequently occurring quantifications of measurement
noise make use of the first and second moment operators of $E$ and $E^{\mm}$.
We will see that these measures of noise have limited applicability, although
they are useful if applied correctly, as shown e.g. in \cite{Oza4}.
It is well known that, in general, a probability measure cannot be determined from its first
and second moments. Therefore, it is natural to expect that a measure of noise or disturbance
based on first and second moments only is equally inadequate.

\subsection{A measure of noise in terms of variances}\label{vari}

We start by analyzing the variance of the probability measure
$E_{\psi }^{\mathcal{M}}$,
$$
\mathrm{Var}\left( E_{\psi }^{\mathcal{M}}\right) = \int \left( x-
\int x\ dE_{\psi }^{\mathcal{M}}(x) \right) ^{2}dE_{\psi
}^{\mathcal{M}}\left( x\right),
$$
which we write as
\begin{eqnarray}
\mathrm{Var}\left( E_{\psi }^{\mathcal{M}}\right) &=&\langle \psi
|\left( E^{\mathcal{M}}\left[ 2\right] -
E^{\mathcal{M}}\left[1\right] ^{2}\right) \psi \rangle  \label{var1}\\
&&\quad +\left( \langle \psi |E^{\mathcal{M}}\left[ 1\right]
^{2}\psi \rangle -\langle \psi |E^{\mathcal{M}}\left[ 1\right]
\psi \rangle ^{2}\right) \notag.
\end{eqnarray}
Here $E^{\mathcal{M}}[k] =\int x^k\,dE^{\mathcal{M}}(x)$ are the
first ($k=1$) and the second ($k=2$) moment operators of
$E^{\mathcal{M}}$. Both terms in the last sum are non-negative,
the first describing the deviation of $E^{\mathcal{M}}$ from being
a projection measure and the second term being the variance of the
spectral measure of the operator $E^{\mathcal{M}}\left[ 1\right]$
in the state $\psi$. The first term is  zero for all $\psi$
exactly when $E^{\mathcal{M}}$ is a projection measure, see e.g.
\cite[Appendix, Sect. 3]{Riesz}. Thus, among the positive operator
measures having the selfadjoint operator
$C:=E^{\mathcal{M}}\left[ 1\right]$ as their first moment
operator, the spectral measure $E^{C}$ has the least variance,
that is, $\mathrm{Var(}E_{\psi }^{\mathcal{M}})\geq
\mathrm{Var(}E_{\psi}^{C})$ for all $\psi$.

Assume now that the measurement process $\mathcal{M}$ (see
Appendix A for technical details) is intended to measure an
observable given by a spectral measure $E=E^{A}$. If $\mathcal{M}$
is \emph{unbiased}, that is, $E^{\mathcal{M}}\left[ 1\right] =A$,
then eq. (\ref{var1}) gives:
\begin{equation*}
\mathrm{Var}\left( E_{\psi }^{\mathcal{M}}\right) = \langle \psi
|\left( E^{\mathcal{M}}\left[ 2\right] -A^{2}\right) \psi \rangle
+\mathrm{Var}\left( E_{\psi }^{A}\right)
\end{equation*}
The positivity of the operator
$N\left(E^{\mathcal{M}},A\right):=E^{\mathcal{M}} \left[ 2\right]
-A^{2}$ suggests to define the number
\begin{equation}\label{noise1}
\epsilon_n(E^{\mathcal{M}},A,\psi)=\ip{\psi}{N(E^{\mm},A)\psi}^{\frac{1}{2}}
\end{equation}
as a quantification of the imprecision of the measurement
$\mathcal{M}$ as a measurement of $A$. With this noise concept one
may write
\begin{equation}\label{var2}
\mathrm{Var}\left( E_{\psi }^{\mathcal{M}}\right)
=\mathrm{Var}\left( E_{\psi }^{A}\right) +
\epsilon_n(E^{\mathcal{M}},A,\psi)^2.
\end{equation}
We thus see that two of the listed criteria for measurement noise
are satisfied: $\epsilon_n(E^{\mathcal{M}},A,\psi)$ is a function
of the probability measures $E^{\mathcal{M}}_\psi$ and $E^A_\psi$,
and this function vanishes when the probability measures are
identical.

This analysis is well-known and it essentially appears already in
one of the earliest monographs on quantum information theory, a
book preprint by R. Ingarden from 1974 \cite{Ing}.

An unbiased measurement $\mathcal M$ of $A$ is  noiseless in a
state $\psi$, that is,  $\epsilon_n(E^{\mathcal{M}},A,\psi)=0$,
exactly when the variances $\mathrm{Var}\left( E_{\psi
}^{\mathcal{M}}\right)$ and $\mathrm{Var}\left( E_{\psi
}^{A}\right)$ are the same. Since for any $\psi\in\mh$,
$\epsilon_n(E^{\mathcal{M}},A,\psi)
=\no{N(E^{\mm},A)^{\frac{1}{2}}\psi}$, we have
$$
\epsilon_n(E^{\mathcal{M}},A,\psi)=0\ \iff\
E^{\mm}[2]\psi=A^2\psi.
$$
Therefore, we also have
\begin{eqnarray*}
&& \epsilon_n(E^{\mathcal{M}},A,\psi)=0 \ {\rm for\ all}\  \psi \ \iff\  \\
&& E^{\mathcal{M}}_\psi=E^{A}_\psi \ {\rm for\ all}\  \psi, \ {\rm
that\ is}\ E^{\mathcal{M}}=E^{A}.
\end{eqnarray*}

The remaining criterion demands that vanishing noise
$\epsilon_n(E^{\mathcal{M}},A,\psi)=0$ should imply the equality
of the probability measures $E^{\mathcal{M}}_\psi$ and
$E^{A}_\psi$. We do not know if this is satisfied by
$\epsilon_n(E^{\mathcal{M}},A,\psi)$. The following two examples
show cases where the quantity defined in
(\ref{noise1}) seems to be a natural measure of noise (Example
\ref{approx}), and where this noise concept may appear misleading
(Example \ref{can}).

\begin{example}\label{approx}
Let $Q$ be the ordinary position observable with the spectral measure
$E^Q:{\mathcal B}(\R)\to L^2(\R)$ and let $f$ be a probability density. The
formula $X\mapsto (\chi_X*f)(Q)=:Q_f(X)$, where $\chi_X*f$ is the convolution
of the characteristic function $\chi_X$ with $f$, defines an approximate
position observable $Q_f$, and one finds that $f$ defines a confidence measure
describing the inaccuracy involved in the $Q$-measurement, see e.g.
\cite[Sect. 3.3]{Davies}. In this case the noise is state
independent.  In fact, for any $\psi$,
$\epsilon_n(Q_f,Q,\psi)=\mathrm{Var}(f)>0$. Here small noise indicates a fairly
accurate position measurement. A measurement model analysis of this well-known
example can be found, for instance, in \cite{OQP},
and it can be traced back to \cite[Sect. VI.3]{vN}.
\end{example}

\begin{example}\label{can}
The canonical phase observable $E^{\rm can}$ with its first moment operator
$\Phi$ gives an example where the noise $\epsilon_n(E^{\rm can},\Phi,\psi)$ can
be made arbitrarily small with an appropriate choice of $\psi$. It can
be argued that this does not indicate that the actual $E^{\rm
  can}$-measurement is an accurate $\Phi$-measurement. Though the
spectrum of $\Phi$ is the phase interval $[0,2\pi)$, the sharp observable $\Phi$
is not a phase observable since it is not covariant under the shifts generated by the number observable. (For
  a recent overview of the theory of covariant phase observables, see
  e.g. \cite{JP}.) That $\epsilon_n(E^{\rm can},\Phi,\psi)$ can be
  made small is due to the fact that $E^{\rm can}$ has the
  norm-1-property, that is, for any $X$ of nonzero Lebesgue measure,
  $\no{E^{\rm can}(X)}=1$ \cite{norm1}. This property implies that the
  variance $\mathrm{Var}(E^{\rm can}_\psi)$ can be made arbitrarily
  small
 \cite[Prop. 2]{norm1}~. From equation (\ref{var2}) it
  is clear that when $\mathrm{Var}(E^{\rm can}_\psi)$ approaches zero,
  also $\mathrm{Var}(E^{\Phi}_\psi)$ and $\epsilon_n(E^{\rm
    can},\Phi,\psi)$ are approaching zero. It is an open question
  whether $\epsilon_n(E^{\rm can},\Phi,\psi)=0$ for some vector state $\psi$.
\end{example}

\subsection{Measurement noise in terms of the difference of two operators}\label{square}

A somewhat different approach to defining the measurement noise in
an approximate measurement of $A$ in a state $\psi$ by means of a
scheme $\mathcal{M}=\langle\mk,\xi,E^M,U\rangle$ (see
Appendix~\ref{A1}) was taken by Ozawa \cite{Oza1, Oza3, Oza5}:
\begin{equation}
\epsilon \left( A,\psi ,\mathcal{M}\right) ^{2}=\langle \psi
\otimes \xi |\left( M^{\mathrm{out}}-A^{\mathrm{in}}\right)
^{2}\psi \otimes \xi \rangle \label{eps1}.
\end{equation}
Here $M^{\mathrm{out}}=U^{\ast }I\otimes MU$ and $A^{\mathrm{in}}=A\otimes I$.
(This characterization of noise is used frequently, for instance, in quantum
optics, see, e.g., \cite{anna1} or \cite{anna2}.) For the sake of comparison we
write the noise $\epsilon_n(E^{\mathcal{M}},A,\psi)$ with the same notations:
\begin{equation}
\epsilon_n(A,\psi,\mathcal M)^2
=\langle \psi \otimes \xi |\left( (M^{\mathrm{out}})^2-
(A^{\mathrm{in}})^2\right)\psi \otimes \xi \rangle \label{var3}.
\end{equation}
We stress that in contrast to (\ref{var3}), in (\ref{eps1}) it is
not assumed that the measurement is unbiased. If the condition
$E^{\mathcal{M}}\left[ 1\right] =A$ is fulfilled, then we have
\begin{equation*}
\epsilon \left( A,\psi ,\mathcal{M}\right)^{2}=
\epsilon_n(A,\psi,\mathcal M)^2
= \mathrm{Var}\left( E_{\psi }^{\mathcal{M}}\right) -
\mathrm{Var}\left( E_{\psi }^{A}\right)
\end{equation*}
and these two notions of noise coincide.

In Appendix~\ref{A1} it will be shown that the quantity
(\ref{eps1}) can be written as:
\begin{equation}\label{eps2}
\epsilon \left( A,\psi ,\mathcal{M}\right)^{2}=\langle \psi
|\left( E^{\mathcal{M}}\left[ 2\right] -E^{\mathcal{M}}\left[
1\right] ^{2}\right) \psi \rangle +\langle \psi |\left(
E^{\mathcal{M}}\left[ 1\right] -A\right)^{2}\psi \rangle
\end{equation}
Thus, $\epsilon(A,\psi,\mm)$ is a function of $A,\psi$ and $E^{\mm}$.
Each of the terms in eq.~(\ref{var2}), or (\ref{var3}), has a simple
operational meaning in that it can be obtained from the statistics of
measurements of $E^{\mathcal{M}}$ and $E^{A}$, performed on two separate
ensembles in the state $\psi$. By contrast, this is not true in general for eq.
(\ref{eps2}): there the second summand contains the operator
$E^{\mathcal{M}}\left[ 1\right] -A$, which cannot be measured together with
$E^{\mathcal{M}}\left[ 1\right]$ or $A$ if these operators do not commute with
respect to $\psi$. In that case, a measurement of the selfadjoint operator
$E^{\mathcal{M}}\left[ 1\right] A+AE^{\mathcal{M}}\left[ 1\right]$, which
occurs in
\begin{equation*}
\begin{split}
\langle \psi |\left( E^{\mathcal{M}}\left[ 1\right] -A\right)
^{2}\psi\rangle & = \langle \psi |E^{\mathcal{M}}\left[ 1\right]
^{2}\psi \rangle +\langle \psi |A^{2}\psi \rangle \\
&\quad -\langle \psi |\left( E^{\mathcal{M}}\left[ 1\right]
A+AE^{\mathcal{M}}\left[ 1\right] \right) \psi \rangle,
\end{split}
\end{equation*}
will in general require a process that cannot be reduced to measurements of
$E^{\mathcal{M}}$ and $A$. In view of eq.~(189) of \cite{Oza5} we note that the
expectation value $\langle \psi |\left( E^{\mathcal{M}}\left[ 1\right]
A+AE^{\mathcal{M}}\left[ 1\right] \right) \psi \rangle$ may be written as a
combination of the expectation values of $E^{\mathcal{M}}\left[ 1\right]$ in
the (nonnormalized) vector states $\psi$, $A\psi$, and $(A+I)\psi$. This is
just another way of expressing the fact that the measurement of the number
$\langle \psi |\left( E^{\mathcal{M}}\left[ 1\right] A+AE^{\mathcal{M}}\left[
1\right] \right) \psi \rangle$ cannot be achieved by measuring $A$ and
$E^{\mathcal{M}}\left[ 1\right]$ in the state $\psi$ only. This state of matter
is also demonstrated in Example \ref{spin1} below.

From eq. (\ref{eps2}) it follows that
$$
\epsilon \left( A,\psi ,\mathcal{M}\right)=0\ \iff\
E^{\mathcal{M}}\left[ 2\right] \psi =E^{\mathcal{M} }\left[
1\right]^{2}\psi\ \& \ E^{\mathcal{M}}\left[ 1\right] \psi =A\psi.
$$
Therefore, as claimed in \cite{Oza1,Oza3}, the following conditions are
equivalent:
\begin{itemize}
\item[(a)]
$\epsilon \left( A,\psi ,\mathcal{M}\right) =0$ for all $\psi$;
\item[(b)] $E^{\mathcal{M}}_\psi=E^{A}_\psi$ for all  $\psi$, that is,  $E^{\mathcal{M}}=E^{A}$.
\end{itemize}
(In \cite{Oza5} Ozawa gives a different proof for this result.) On the basis of
this result one may ask if for a given $\psi$ the condition $\epsilon \left(
A,\psi ,\mathcal{M}\right) =0$ still is equivalent with
$E^{\mathcal{M}}_\psi=E^{A}_\psi$. Example \ref{spin3} shows that one can have
$\epsilon \left( A,\psi ,\mathcal{M}\right) = 0$ without the probability
measures $E^{\mathcal{M}}_\psi$ and $E^{A}_\psi$ being equal. If $\epsilon
\left( A,\psi ,\mathcal{M}\right) = 0$, then the first and second moments of
the probability measures $E^{\mathcal{M}}_\psi$ and $E^{A}_\psi$ are the same.
On the other hand, even equality of all moments does not guarantee that the
noise is zero. Indeed, examples \ref{spin2} and \ref{QP} show that the
probability measures $E^{\mathcal{M}}_\psi$ and $E^{A}_\psi$ can be the same
although $\epsilon \left( A,\psi ,\mathcal{M}\right) \ne 0$.

In the special case of $E^{\mathcal{M}}$ being a spectral measure
$E^{C}$ eq. (\ref{eps2}) takes the form
\begin{equation}\label{epsC}
\epsilon \left( A,\psi ,\mathcal{M}\right) ^{2}=\langle \psi
|(C-A)^{2}\psi \rangle
\end{equation}
and $\epsilon \left( A,\psi ,\mathcal{M}\right)=0$ exactly when
$A\psi=C\psi$.

\begin{example}\label{spin1}
Assume that one intends to measure the component $A=s_{\vec{a}}$
of the spin of a spin-$\frac{1}{2}$ object. Assume also that there is
a systematic error in the
measurement (e.g. misalignment of the magnet) meaning
that one is actually measuring some
component $C=s_{\vec{c}}$, with $\vec{c}$ a unit vector close to
$\vec{a}$. Then, for any vector state $\psi$ we get
\begin{equation*}
\epsilon \left(s_{\vec{a}},\psi ,\mathcal{M}\right) ^{2}=\langle
\psi |(s_{\vec{c}}-s_{\vec{a}})^{2}\psi \rangle = \frac
12(1-\vec{c}\cdot\vec{a}).
\end{equation*}
Clearly, $\epsilon \left(s_{\vec{a}},\psi ,\mathcal{M}\right)$ tends to zero
with $\vec{c}\cdot\vec{a}$ approaching 1, but the operator
$s_{\vec{c}}-s_{\vec{a}}$  does not commute with $s_{\vec{c}}$ or
$s_{\vec{a}}$. Actually all these operators are pairwisely totally
noncommutative, unless $\vec{c}\cdot\vec{a}=\pm 1$. An estimate of $\epsilon
\left(s_{\vec{a}},\psi ,\mathcal{M}\right)$ cannot therefore be obtained from
the statistics of measurements of $s_{\vec{a}}$ and $s_{\vec{c}}$ in
the state $\psi$ only. To estimate $\epsilon
\left(s_{\vec{a}},\psi ,\mathcal{M}\right)$ one should either do
measurements in other states than $\psi$ or measure some other
observables than $s_{\vec{a}}$ and $s_{\vec{c}}$.
\end{example}

\begin{example}\label{spin2}
Continuing with Example \ref{spin1}, assume that the system is in a spin state
$\psi_{\vec{n}}$, a $\frac 12$\,-eigenstate of a spin component $s_{\vec{n}}$.
Then
$\ip{\psi_{\vec{n}}}{s_{\vec{a}}\psi_{\vec{n}}}=\frac{1}{2}\vec{n}\cdot\vec{a}$ and
$\ip{\psi_{\vec{n}}}{s_{\vec{c}}\psi_{\vec{n}}}=\frac{1}{2}\vec{n}\cdot\vec{c}$
showing that
the spin observables $s_{\vec{a}}$ and $s_{\vec{c}}$ have same
probabilities in the state $\psi_{\vec{n}}$ exactly when
$\vec{n}\cdot\vec{a}=\vec{n}\cdot\vec{c}$, i.e., when the angle between $\vec{n}$ and
$\vec{a}$ is the same as the angle between $\vec{n}$ and $\vec{c}$.
Thus, it may happen that  the probability distributions for
$s_{\vec{a}}$ and   $s_{\vec{c}}$
in a given state $\psi_{\vec{n}}$ are the
same, but the noise $\epsilon \left(s_{\vec{a}},\psi ,\mathcal{M}\right)$ is nonzero.
\end{example}

In formula (\ref{epsC}) no restrictions are given for the selfadjoint operators $A$ and $C$, except that $C$
is obtained by the measurement process $\mathcal M$.
Therefore, its blind application may lead to  unexpected or  unwanted results.
This is demonstrated by
Examples \ref{spin3} and \ref{QP}, which indicate that the actually measured quantity, here $C$, should
somehow be related with the quantity which is intended to be measured, here $A$.
\begin{example}\label{spin3}
Consider two selfadjoint matrices $A$ and $C$ in $\C^2$,
$$
A= \frac{1}{2}\left( \begin{array}{cc} 1 & 0 \\
0 & -1
\end{array}\right),
\quad C= \frac{1}{8} \left( \begin{array}{cc} 3 & 5 \\
5 & 3
\end{array}\right).
$$
If $\psi=\frac{1}{\sqrt{10}} (-3,1)^T$, then $A\psi=C\psi$, which means that
$\epsilon(A,\psi,\mm)=\ip{\psi}{(A-C)^2\psi}=0$,
though the probability distributions are different.
Clearly, matrices $A$ and $C$
have  different eigenvalues but also all the probabilities in the state $\psi$
are different.
\end{example}

\begin{example}\label{QP}
Let now $A=Q$ and $C=P$ be  the usual multiplicative (position) and
 differential (momentum) operators acting in  the Hilbert space $L^2(\R)$.
In this case,  for all $\psi\in L^2(\R)$,
$\epsilon(Q,\psi,\mm)
\neq 0$.
However, if a function $\psi$ is identical with its Fourier transform, then
the probability distributions  $E^Q_\psi$ and $E^P_\psi$ are the same.
\end{example}

Though artificial,
Examples \ref{spin3} and \ref{QP} seem to suggest that in order to apply the
quantity (\ref{eps1}) as a measure of noise in a measurement $\mathcal M$ of
$A$, some further restrictions on $\mathcal M$ have to be posed, as is the
case, for instance, in Example \ref{approx}.

The quantity $\epsilon \left( A,\psi ,\mathcal{M}\right)$ is mathematically
well-defined and it has the important property that $\epsilon \left( A,\psi
,\mathcal{M}\right)=0$ for all $\psi$ if and only if $E^{\mm}=E^A$. However,
its interpretation as a measure of noise in measuring $A$ in the state $\psi$
with the scheme $\mathcal M$ seems to require either that $\mm$ is unbiased or
that $A$ and $E^{\mathcal{M}}$ are jointly measurable in the state $\psi$.
Furthermore, it is not obvious how this measure of noise should be adapted to
observables $E$ which cannot be represented as selfadjoint operators (like
covariant phase observables). These observations lead back to the original
question of finding a quantitative, operationally meaningful, measure of the
difference between $E^{\mathcal{M}}$ and $E$ where these positive operator
measures are actually different and non-coexistent (in the sense of Ludwig
\cite{Ludwig}).

\subsection{Measurement noise and the total variation norm}\label{total}

In order to compare two operator measures, one usually needs to
compare all their moment operators. In the case of bounded
operator measures, equality of all moment operators guarantees the
equality of the operator measures. However, it is well-known that
there are pairs of unbounded measures for which all the moment
operators coincide but the measures are different \cite{momprob}.
In either case it is clear that one cannot expect that any
quantity composed of first and second moments only would be
sufficient to characterize the difference of two operator
measures.

Quantum mechanics is a statistical theory and measurements give probability
distributions. The most obvious way to estimate the difference of quantum
observables seems to be the comparison of their probability distributions. This
can be done by choosing a metric or a norm in a set of probability measures.
One example is the total variation norm $\no{\cdot}_1$. We recall that for a
measure $\mu$ the total variation norm is defined as $\left\|\mu \right\|_1:=
\sup\sum_1^n|\mu(X_k)|$ where the supremum is taken over $(X_k)_1^n$ finite
partitions of $\mathbb R$. Clearly, the number $\left\| E_{\psi
}^{\mathcal{M}}-E_{\psi }\right\|_1$ can be obtained from the measurement
outcome statistics of the observables in question and therefore the total
variation norm is operationally meaningful. Now one has for each vector state
$\psi$:
\begin{equation*}
\left\| E_{\psi }^{\mathcal{M}}-E_{\psi }\right\|_1 =0\;\; \;\;\;\iff
\;\;\;E_{\psi}^{\mathcal{M}}=E_{\psi}.
\end{equation*}
This also implies that
\begin{equation*}
\left\| E_{\psi }^{\mathcal{M}}-E_{\psi }\right\|_1
=0\;\;\mathrm{for\;all\;} \psi \;\;\;\iff \;\;\;E^{\mathcal{M}}=E.
\end{equation*}

Though the total variation norm has a clear operational meaning it
does not seem to lend itself easily to quantify the intuitive idea
on measurement inaccuracy or disturbance expressed in (UP1).

\subsection{The quantity $\langle \psi |(AE^{\mm}[1]+E^{\mm}[1]A)\psi \rangle$ and covariance}

In Section \ref{square} we saw that the noise $\epsilon(A,\psi,\mm)$
contains a term
\begin{equation*}
\langle \psi |(AE^{\mm}[1]+E^{\mm}[1]A)\psi\rangle
\end{equation*}
and the problem in its operational meaning was pointed out. In some
 cases the number
\begin{equation}\label{AE2}
\frac{1}{2}\langle \psi |(AE^{\mm}[1]+E^{\mm}[1]A)\psi\rangle -
\langle \psi |A\psi\rangle \langle\psi| E^{\mm}[1]\psi\rangle
\end{equation}
gives the covariance of the observables $A$ and $E^{\mm}$ in their
joint measurement. However, we will demonstrate that, in general, this
kind of interpretation is problematic.

\begin{example}\label{cov:QP}
Let $Q$ and $P$ be the ordinary position and momentum operators acting
in $L^2(\mathbb R)$.
These operators are totally noncommutative and therefore the number (\ref{AE2}), with $A=Q$ and $E^{\mm}[1]=P$,
 cannot be interpreted
as their covariance in each state $\psi$.
However, as well-known, there are phase space distributions for which the covariance takes the form (\ref{AE2}).

Let $W_\phi$ be the  Wigner distribution of a Gaussian state  $\phi\in L^2(\mathbb R)$.
It  is a probability density
for which
$$
\mathrm{Cov}(W_\phi;x,y)=\frac{1}{2}\langle \phi |(QP+PQ)\phi\rangle -
\langle \phi |Q\phi\rangle \langle\phi|P\phi\rangle=0.
$$
The Wigner distribution $W_\psi$ of an arbitrary state $\psi$  has the position and momentum
distributions $E^Q_\psi$ and $E^P_\psi$ as the marginal distributions. However, $W_\psi$ is a
probability distribution  only for the Gaussian states \cite{Hudson}  so that, in general, $\mathrm{Cov}(W_\psi;x,y)$
does not have a probabilistic meaning, yielding, thus,  no similar interpretation
for the quantity $\langle \psi |(QP+PQ)\psi\rangle$.

The Husimi distribution $H_\psi$ of any state $\psi\in L^2(\mathbb R)$ is a probability distribution
and for it we get
$$
\mathrm{Cov}(H_\psi;x,y)=\frac{1}{2}\langle \psi |(QP+PQ)\psi\rangle -
\langle \psi |Q\psi\rangle \langle\psi|P\psi\rangle
$$
for any $\psi$ (for which the relevant integrals exist).
The marginal distributions of the Husimi distribution $H_\psi$
are not the position and momentum distributions $E^Q_\psi$ and $E^P_\psi$
being the probability distributions of unsharp position and momentum
observables, compare to Example~ \ref{approx}.
Indeed, $H_\psi$ is the density
of the probability measure $\mu_\psi$ defined by the phase space observable $A^{\ket 0}$
(associated with the oscillator Gaussian ground state $\ket 0$)
and the state $\psi$, and the Cartesian marginal observables of
$A^{\ket 0}$ are the approximate
position and momentum observables  \cite[Sections 3.3 and 3.4]{Davies}.
In this case, therefore, the covariance $\mathrm{Cov}(H_\psi;x,y)$ is the covariance of
approximate position and momentum observables, not of $Q$ and $P$.
\end{example}

\begin{example}
The Husimi distribution $H_\psi$  of Example \ref{cov:QP} gives rise to another example when we
use the polar coordinates $(r,\theta)$. The angle marginal measure of the phase space observable
$A^{\ket{0}}$ is a (phase shift covariant) phase observable
$A^{\ket{0}}_{\theta}$ and the radial marginal measure $A^{\ket{0}}_{r}$ is a
smeared number observable. Their first moment operators are
\begin{eqnarray*}
A^{\ket{0}}_{\theta}[1] &=& \sum_{n\neq m=0}^{\infty} \frac{i\
\Gamma(\frac{n+m}{2}+1)}{\sqrt{n!m!}(m-n)}\ \kb{n}{m}+\pi I, \\
A^{\ket{0}}_r[1] &=& N+I,
\end{eqnarray*}
see, for instance, \cite{LP99} and \cite{Grabowski93}. Thus,
for any oscillator eigen state $\ket n$ one gets
\begin{equation*}
\frac{1}{2}\langle
n|(A^{\ket{0}}_{\theta}[1]A^{\ket{0}}_{r}[1]+A^{\ket{0}}_{r}[1]A^{\ket{0}}_{\theta}[1])|n\rangle
=(n+1)\pi
\end{equation*}
but
\begin{equation*}
\int r\theta\,d\mu_{\ket{n}}= n!\pi,
\end{equation*}
showing that the covariance $\mathrm{Cov}(H_{\ket n};r,\theta)$
cannot be obtained
from an expression of the form (\ref{AE2}).
\end{example}

There are plenty of physically important cases
 where the covariance in the form (\ref{AE2}) and the noise
 (\ref{eps1}) are operationally meaningful. This is especially guaranteed
 whenever the observables $A$ and $E^{\mm}$ commute. Next we discuss
 this situation.

Assume that the observables $A$ and $E^{\mm}$ commute in all states $\psi$.
Then the map
$$
X\times Y\mapsto \langle \psi
|E^A(X)E^{\mathcal M}(Y)\psi\rangle
$$
extends to a probability measure $\mu_\psi$ on $\bor{\R^2}$
 and its (Cartesian) marginal measures are $E_{\psi}^A$
and $E_{\psi}^{\mm}$. One also obtains
\begin{equation*}\label{eps4}
\epsilon \left( A,\psi ,\mathcal{M}\right)^{2} = \int \left(
x-y\right) ^{2}\,d\mu_{\psi}(x,y),
\end{equation*}
and
\begin{equation*}\label{cov3}
\int xy\,d\mu_{\psi} = \frac 12  \ip{\psi}{(A
E^{\mathcal{M}}[1]+E^{\mathcal{M}}[1]A)\psi},
\end{equation*}
so that, in particular, the value of $\epsilon \left(
A,\psi,\mathcal{M}\right)$ can be estimated from the statistics of
a joint measurement of $E^{\mathcal{M}}$ and $A$. We can also
write
\begin{eqnarray}\label{cov4}
\epsilon \left( A,\psi ,\mathcal{M}\right)^{2} &=& \left(
\mathrm{Exp}\left(E_{\psi }^{\mathcal{M}}\right)
 - \mathrm{Exp}\left(E_{\psi }^{A}\right) \right)^2 \\
 &+&  \left(
\sqrt{\mathrm{Var}\left(E_{\psi}^{\mm}\right)}
-
\sqrt{\mathrm{Var}\left( E^A_{\psi}\right)}
\right)^2 \nonumber \\
&+&  2\left( \sqrt{\mathrm{Var}\left( E_{\psi}^{\mm}\right)
\mathrm{Var}\left( E_{\psi
}^{A}\right)}-\mathrm{Cov}(\mu_{\psi}) \right) \nonumber
\end{eqnarray}
showing that higher covariance means lower noise.

The following example, which comes from the class of standard
measurement models \cite{OQP}, demostrates the previous discussion.

\begin{example}
Consider a nondemolition measurement of the photon
number of a single mode optical field, applying  a two-mode
coupling of the form
\begin{equation*}\label{inter1}
U=e^{i\chi N_1\otimes N_2},
\end{equation*}
where $N_1=a_1^*a_1=\sum n_1\kb{n_1}{n_1}$ and $N_2=a_2^*a_2=\sum
n_2\kb{n_2}{n_2}$ are the number observables of the signal mode and the probe
mode, respectively, and $\chi$ is a real coupling constant. Fix an  initial
vector state $\phi$ of the probe mode and choose a probe observable $E^M$ as
the pointer observable. The
measurement scheme, which aims to measure $N_1$, is thus defined by $U,\phi$
and $E^M$. The actually measured observable is a smeared  number
observable $N_1$,
\begin{equation*}
E^{\mm}(X)=\sum_{n=0}^{\infty}\ip{\phi}{e^{-i\chi n
N_2}E^M(X)e^{i\chi n N_2}\phi}\ \kb{n}{n}, \quad X\in\br,
\end{equation*}
so that $N_1$ commutes with $E^{\mm}$. Though $E^{\mm}[1]\ne N_1$ in general,
the moment operators of $E^{\mm}$ are functions of $N_1$,
$$
E^{\mm}[k]= \sum_{n=0}^{\infty}\ip{\phi}{e^{-i\chi n
N_2}M^ke^{i\chi n N_2}\phi}\ \kb{n}{n},\quad k\in\N.
$$
In this case, for any vector state  $\psi$ of the signal mode one
gets
\begin{eqnarray*}
&& \epsilon(N_1,\psi,\mm)=\ip{\psi}{\left(E^{\mm}[2]-2E^{\mm}[1]N_1+N^2_1\right)\psi}, \\
&& \int xy\,d\mu_{\psi}=\ip{\psi}{E^{\mm}[1]N_1\psi},
\end{eqnarray*}
whenever the integrals in question converge and where $\mu_{\psi}$
is the probability measure extending the map $X\times \{n_1\}
\mapsto\ip{\psi}{E^{\mm}(X)\kb{n_1}{n_1}\psi}$.
\end{example}

To conclude, if $E^A$ and $E^{\mm}$ commute, 
 then the covariance and the noise are operationally well-defined and
 they are linked by eq. (\ref{cov4}). However, in general these
 concepts are problematic.

\section{Measurement disturbance}

The initial state of a system will in general change under the
influence of a measurement; there is no (nontrivial) measurement
which would leave unchanged all the states of the system.
If the object system is initially in a
vector state $\psi$, its state after applying the measurement
process $\mathcal M$ is $\mathcal I(\mathbb R)(P[\psi])$. The state $\mathcal I(\mathbb R)(P[\psi])$ is
the unique state of the object system  obtained by tracing out the probe degrees of
freedom from the final object-probe state $U(\psi\otimes\xi)$ (see Appendix \ref{A2} for technical details) .
If $B$ is an arbitrary object observable (a bounded selfadjoint
operator on $\mh$), then under the influence of the measurement
process $\mathcal M$, the measurement outcome probabilities for
$B$ get changed from $E^B_{\psi}$ to $E^B_{\mathcal I(\mathbb
R)(P[\psi])}$. The difference between these probability measures
describes the influence of the measurement of $A$ implemented by
 $\mathcal M$ on the $B$-probabilities.
Alternatively, using the Heisenberg picture, the observable $B$, with the
spectral measure $E^B$, is changed into an observable $E$  defined as
$$
E(X)=\mathcal I(\mathbb R)^*(E^B(X)),
$$
where $\mathcal I(\mathbb R)^*$ is the dual transform  of the state
transformation $\mathcal I(\mathbb R)$. In general, $E$ is a positive operator
measure. Thus, a study of the measurement disturbance may equally well be based
on a comparison of the operator measures $E^B$ and $E$. In this sense it is
clear that a study of the measurement disturbance is completely analogous to a
study of the measurement noise.
We do not repeat all the analysis of Section \ref{mnoise} in this context.
Rather, we shall point out some special aspects of the problem.

The moment operators of $E$ can easily be computed, and one gets
\begin{equation*}
E[1]=\mathcal I(\mathbb R)^*(B),\ \ \ E[2]=\mathcal I(\mathbb R)^*(B^2).
\end{equation*}
We note that if $E[1]=B$, then for any state
$$
{\rm Var}(E,\psi)={\rm Var}(B,\mathcal I(\mathbb R)(P[\psi])) \geq {\rm Var}(B,\psi),
$$
with an equality (for all states) if and only if $E=E^B$, that is, if and only if
$E[2]=B^2$.
It is interesting to remark that  the invariance of the selfadjoint operator $B$
under the measurement, that is, $\mathcal I(\mathbb R)^*(B)=B$, does not guarantee
the invariance of the observable $E^B$ under $\mathcal M$, that is,
the invariance of
$B^2$ under $\mathcal I(\mathbb R)^*$. An
example demonstrating this fact is constructed in
\cite{Gudder}.

In \cite{Oza1, Oza3} it is proposed that the following quantity
serves to describe the {\em disturbance} of the measurement
$\mathcal M$ on $B$, intended to measure $A$:
\begin{equation*}
\eta(B,\psi,A)^2=\ip{\psi\otimes\xi}{\left(B^{\rm out} - B^{\rm
in}\right)^2\psi\otimes\xi}\label{eta1}.
\end{equation*}
Here, again, $B^{\rm out} =U^*B\otimes IU$ and $B^{\rm
in}=B\otimes I$. In Appendix~\ref{A2} it will be shown that this
quantity can be expressed in the form:
\begin{eqnarray*}
\eta(B,\psi,A)^2
&=& \ip{\psi}{\left(\mathcal I(\mathbb R)^*(B^2)-(\mathcal I(\mathbb R)^*(B))^2\right)\psi}\label{eta2}\\
&&\quad +\ip{\psi}{\left(\mathcal I(\mathbb
R)^*(B)-B\right)^2\psi}\notag\\
&=&\ip{\psi}{(E[2]-E[1]^2)\psi}\notag\\
&&\quad
+\ip{\psi}{\left(E[1]-B\right)^2\psi}.\notag
\end{eqnarray*}
Since the operators $E[2]-E[1]^2$ and $(E[1]-B)^2$ are positive
 we obtain that $\eta(B,\psi,A)=0$ exactly when
$\mathcal I(\mathbb R)^*(B)\psi=B\psi$, i.e. $E[1]\psi=B\psi$,
 and $\mathcal I(\mathbb
R)^*(B^2)\psi=\mathcal I(\mathbb R)^*(B)^2\psi$,
i.e. $E[2]\psi=E[1]^2\psi$.
Thus we come to
the following result:
\begin{equation*}\label{eta3}
\eta(B,\psi,A) = 0 {\rm\ for\ all\ } \psi \ \iff\ \mathcal
I(\mathbb R)^*(B) = B\  \& \ \mathcal I(\mathbb R)^*(B^2)= B^2,
\end{equation*}
that is,
 \begin{equation*}
\eta(B,\psi,A) = 0 {\rm\ for\ all\ } \psi \ \iff\
E=E^B.
\end{equation*}
(This result was stated in \cite{Oza1,Oza3} and proved by different
methods in the preprint \cite{Oza5}.)

The measurement interaction is modelled by a unitary operator $U$. Therefore,
the map  $\mathcal I(\mathbb R)^*$ is completely positive so that
 there is a sequence of bounded operators $D_i$ such that $\mathcal
I(\mathbb R)^*(B) =\sum D_i^*BD_i$ (for all $B$, convergence
ultraweakly). Moreover, since  $\mathcal I(\mathbb R)^*(I)=I$, we have $\sum
D_i^*D_i=I$, see e.g. \cite[Theorem 2.3]{Davies}.
From \cite[Cor. 3.4]{Gudder} it follows that
\begin{equation*}
\mathcal I(\mathbb R)^*(B) = B\  \& \ \mathcal I(\mathbb R)^*(B^2)= B^2\ \iff\
BD_i=D_iB\textrm{ for all $i$}.
\end{equation*}

Hence, the following conditions are equivalent:
\begin{itemize}
\item[(a)] $\eta(B,\psi,A)=0$ for all $\psi$;
\item[(b)] $\mathcal I(\mathbb R)^*(B) = B$ and $\mathcal I(\mathbb R)^*(B^2)=
B^2$;
\item[(c)] $\mathcal I(\mathbb R)^*(E^B(X)) = E^B(X)$ for all
$X\in\bor{X}$;
\item[(d)] $BD_i=D_iB$ for all $i$.
\end{itemize}

When $\eta(B,\psi,A)\ne 0$ there is no guarantee that $E[1]$ and $B$ would
commute, and, therefore, as in the case  of  eq. (\ref{eta2}), the operational
meaning of the quantity $\eta(B,\psi,A)$ remains problematic, being, perhaps,
only of  limited validity.

\begin{remark}
If the A-measurement $\mathcal M$ is noiseless so that $E^{\mathcal M}=E^A$,
then the `distorted observable' $E$, with $E(X)=\mathcal I(\R)^*(E^B(X))$,
always commutes with $A$, showing that a noiseless measurement exhibits a kind
of maximal disturbance. This follows from the fact that the operator bimeasure
$(Y,X)\mapsto \mathcal I(Y)^*(E^B(X))$ extends uniquely to a normalized POM
having $E^A$ and $E$ as its Cartesian marginal measures, see, e.g. \cite{PLKY}
For instance, any noiseless position measurement distorts the conjugate
momentum such that the `distorted momentum' commutes with the position.
\end{remark}

\section{Conclusion}

Each of the three measures of noise (or disturbance) investigated in this paper
have their own merits and shortcomings. Therefore, the limited range of their
applicability must be acknowledged. The problem of quantifying the noise and the disturbance in quantum
measurements remains thus  an important open problem.

\appendix

\section{Proof of Equation \ref{eps2}}\label{A1}

Let $A$ be a bounded selfadjoint operator and consider  a
measurement process $\mathcal{M}= \langle\mk,\xi,M,U\rangle$
planned out to measure $A$. Here $\mk$ is the probe Hilbert space,
$\xi\in\mk$, $\no{\xi}=1$, the initial vector state of the probe,
$M$ the pointer observable, a bounded selfadjoint operator on
$\mk$, and $U:\mh\otimes\mk\to\mh\otimes\mk$ a unitary mapping
modeling the measurement coupling. The actually measured
observable $E^{\mathcal M}$ is uniquely determined by the
probability reproducibility condition \cite{Ozawa84,QTM}
$$
\ip{\psi}{E^{\mathcal M}(X)\psi}
=\ip{\psi\otimes\xi}{U^*I\otimes E^{M}(X)U\,\psi\otimes\xi},
$$
for all $X\in\br,\psi\in\mh$. Since $M$ is assumed to be
bounded, the first and the second moment operators $E^{\mathcal
M}[1]$ and $E^{\mathcal M}[2]$ of $E^{\mathcal M}$ are the bounded
selfadjoint operators for which for all $\psi\in\mh$
\begin{eqnarray*}
\ip{\psi}{E^{\mathcal M}[1]\psi} &=& \ip{\psi\otimes\xi}{U^*I\otimes MU\,\psi\otimes\xi},\\
\ip{\psi}{E^{\mathcal M}[2]\psi}
&=&\ip{\psi\otimes\xi}{U^*I\otimes M^2U\,\psi\otimes\xi}.
\end{eqnarray*}

Consider now the quantity
\begin{equation*}
\epsilon \left( A,\psi ,\mathcal{M}\right)^{2}=\langle \psi
\otimes \xi |\left( M^{\mathrm{out}}-A^{\mathrm{in}}\right)
^{2}\psi \otimes \xi \rangle,
\end{equation*}
where $M^{\mathrm{out}}=U^{\ast }I\otimes MU$ and
$A^{\mathrm{in}}=A\otimes I$. Now
\begin{eqnarray*}
&& \langle \psi \otimes \xi|\,(M^{\mathrm{out}})^{2}\,\psi \otimes
\xi \rangle
=\ip{\psi}{E^{\mathcal M}[2]\psi},\\
&& \langle \psi \otimes \xi|\,(A^{\mathrm{in}})^{2}\,\psi \otimes
\xi \rangle = \ip{\psi}{A^2\psi}.
\end{eqnarray*}
Since $A^{\mathrm{in}}$ commutes with $I\otimes P[\xi]$ and
$$
I\otimes P[\xi] M^{\mathrm{out}} I\otimes P[\xi] =E^{\mathcal
M}[1]\otimes P[\xi],
$$
we also have
\begin{eqnarray*}
&& \langle \psi \otimes \xi | M^{\mathrm{out}}A^{\mathrm{in}}\psi
\otimes \xi \rangle
= \ip{\psi}{E^{\mathcal M}[1]A\psi},\\
&& \langle \psi \otimes \xi | A^{\mathrm{in}}M^{\mathrm{out}}\psi
\otimes \xi \rangle = \ip{\psi}{AE^{\mathcal M}[1]\psi}.
\end{eqnarray*}
Therefore, we get:
\begin{equation*}
\epsilon \left( A,\psi ,\mathcal{M}\right) ^{2} =\langle \psi
|\left( E^{\mathcal{M}}\left[ 2\right] -E^{\mathcal{M}}\left[
1\right] ^{2}\right) \psi \rangle +\langle \psi |\left(
E^{\mathcal{M}}\left[ 1\right] -A\right) ^{2}\psi \rangle.
\end{equation*}

Both terms in the right hand side of this equation are
nonnegative, the first one due to $E^{\mathcal{M}}\left[ 2\right]
\geq E^{\mathcal{M}}\left[ 1\right] ^{2}$, see e.g. \cite{Riesz}.
Therefore, $\epsilon \left( A,\psi ,\mathcal{M}\right)=0$ if and
only if $E^{\mathcal{M}}\left[ 2\right]\psi =E^{\mathcal{M}}\left[
1\right] ^{2}\psi$ and $E^{\mathcal{M}}\left[ 1\right]\psi=A\psi$.
Consequently, since $A$ and $M$ are assumed to be bounded
operators, one gets that $\epsilon \left( A,\psi
,\mathcal{M}\right)=0$ for all $\psi$ exactly when
$E^{\mathcal{M}}$ is a spectral measure and $E^{\mathcal{M}}=E^A$.

We close this appendix with a characterization of
$E^{\mathcal{M}}$ being a spectral measure (not necessarily equal
to $E^A$) in terms of  the measurement scheme $\mathcal{M}$. This
is an immediate consequence of the well-known fact that for any
two projection operators $P$ and $R$, the product $PRP$ is a
projection if and only if $PR=RP$.

\begin{lemma}
The positive operator measure $E^{\mathcal M}$ is a spectral
measure
if and only if the projection operators $I\otimes P[\xi]$  and
$U^*I\otimes E^{M}(X)U$ commute for all $X$.
\end{lemma}

\section{Proof of Equation \ref{eta2}}\label{A2}

Consider the measurement scheme $\mathcal{M}=
\langle\mk,\xi,M,U\rangle$ as introduced in Appendix~\ref{A1}. If
$\psi$ is the initial vector state of the system, then its state
after the measurement $\mathcal M$ is $\mathcal I(\mathbb
R)(P[\psi])$. This is the unique state (positive trace one
operator on $\mh$) for which
\begin{eqnarray*}
{\rm tr}\,[\mathcal I(\mathbb
R)(P[\psi])B] &=& \ip{\psi\otimes\xi}{U^*B\otimes
  E^M(\R)\,U\psi\otimes\xi}\\
&=& \ip{\psi\otimes\xi}{U^*B\otimes I\,U\psi\otimes\xi}
\end{eqnarray*}
for any bounded selfadjoint operator $B$ acting on $\mh$. Using
the dual transformation $\mathcal I(\mathbb R)^*$,
the expression ${\rm tr}\,[\mathcal I(\mathbb R)(P[\psi])B]$ can be written as
${\rm tr}\,[P[\psi]\mathcal I(\mathbb R)^*(B)]=\ip{\psi}{\mathcal I(\mathbb
R)^*(B)\psi}$. It follows that
$$
I\otimes P[\xi] \, B^{\mathrm{out}}\, I\otimes P[\xi] =\mathcal
I(\mathbb R)^*(B) \otimes P[\xi].
$$
Hence,
\begin{eqnarray*}
\eta(B,\psi,A)^2 &=& \ip{\psi\otimes\xi}{(B^{\rm out}-B^{\rm in})^2\psi\otimes\xi}\\
&=&\ip{\psi\otimes\xi}{(B^{\rm out})^2\psi\otimes\xi}+\ip{\psi\otimes\xi}{(B^{\rm in})^2\psi\otimes\xi}\\
&&\quad -2\mathrm{Re}\ip{\psi\otimes\xi}{B^{\rm out}B^{\rm in}\psi\otimes\xi}\\
&=&\ip{\psi}{\mathcal I(\mathbb R)^*(B^2)\psi}+\ip{\psi}{B^2\psi}
-2\mathrm{Re}\ip{\psi}{B\mathcal I(\mathbb R)^*(B)\psi}\\
&=& \ip{\psi}{\left(\mathcal I(\mathbb R)^*(B^2)-(\mathcal I(\mathbb R)^*(B))^2\right)\psi} \\
&&\qquad +\ip{\psi}{\left(\mathcal I(\mathbb
R)^*(B)-B\right)^2\psi}.
\end{eqnarray*}

We give here an alternative proof for the fact that
$\mathcal I(\mathbb R)^*(B^2)\geq (\mathcal I(\mathbb R)^*(B))^2$
using the complete positivity of $\mathcal I(\mathbb R)^*$ with the
representation $\mathcal I(\mathbb R)^*(\cdot)=\sum D_i^*\,\cdot\,D_i$.
Applying twice the Cauchy-Schwarz
inequality one gets for each vector $\psi$:
\begin{eqnarray*}
\no{\mathcal I(\mathbb R)^*(B)\psi}^2
&&= \ip{\mathcal I(\mathbb R)^*(B)\psi}{\mathcal I(\mathbb R)^*(B)\psi}\\
&&=\sum\ip{BD_i\psi}{D_i\mathcal I(\mathbb R)^*(B)\psi}\\
&&\leq\sum\no{BD_i\psi}\no{D_i\mathcal I(\mathbb R)^*(B)\psi}\\
&&\leq\left(\sum\no{BD_i\psi}^2\right)^{1/2}\left(\sum\no{D_i\mathcal I(\mathbb R)^*(B)\psi}^2\right)^{1/2}\\
&&= \ip{\psi}{\mathcal I(\mathbb R)^*(B^2)\psi}^{1/2} \no{\mathcal
I(\mathbb R)^*(B)\psi}.
\end{eqnarray*}
Therefore, for any $\psi\in\mh$, $\ip{\psi}{(\mathcal I(\mathbb
R)^*(B))^2\psi}\leq\ip{\psi}{\mathcal I(\mathbb R)^*(B^2)\psi}$,
that is,  $\mathcal I(\mathbb R)^*(B^2)\geq (\mathcal I(\mathbb
R)^*(B))^2$.
\

\

\noindent
{\bf Acknowledgement.}
The authors wish to thank Dr. Masanao Ozawa for his comments on an earlier (August, 2003) version of this paper.

\end{document}